\documentclass[journal=jacsat,manuscript=article]{achemso}
\usepackage{chemformula} 
\usepackage[T1]{fontenc} 
\usepackage{graphicx}
\usepackage{array}
\usepackage[colorlinks=true,linkcolor=blue,citecolor=blue,urlcolor=blue]{hyperref}
\usepackage{natbib}
\usepackage{tabularx}
\usepackage{multirow}
\usepackage{dcolumn}
\usepackage{mciteplus}

\usepackage{amsmath}

\usepackage{amssymb}

\usepackage{color}
\usepackage{lineno}

 \title{Optical markers of magnetic phase transition in CrSBr}

\author{W.~M.~Linhart}
\email{wojciech.linhart@pwr.edu.pl}
\author{M. Rybak}
\affiliation{Department of Semiconductor Materials Engineering, Faculty of Fundamental Problems of Technology,
 Wroc{\l}aw University of Science and Technology, Wybrze{\.z}e Wyspia{\'n}skiego 27, 50-370 Wroc{\l}aw, Poland}
 \author{M. Birowska}
\affiliation{Institute of Theoretical Physics, Faculty of Physics, University of Warsaw, Pasteura 5, 02-093 Warsaw, Poland}
 \author{K. Mosina}
\affiliation{Department of Inorganic Chemistry, University of Chemistry and Technology Prague, Technick{\'a} 5, 166 28 Prague 6, Czech Republic}
 \author{V. Mazanek}
\affiliation{Department of Inorganic Chemistry, University of Chemistry and Technology Prague, Technick{\'a} 5, 166 28 Prague 6, Czech Republic}
\author{P. Scharoch}
\affiliation{Department of Semiconductor Materials Engineering, Faculty of Fundamental Problems of Technology,
 Wroc{\l}aw University of Science and Technology, Wybrze{\.z}e Wyspia{\'n}skiego 27, 50-370 Wroc{\l}aw, Poland}

\author{D. Kaczorowski}
\affiliation{Institute of Low Temperature and Structure Research, Polish Academy of Sciences, ul. Ok{\'o}lna 2, 50-422 Wroc{\l}aw, Poland}
\author{Z. Sofer}
\affiliation{Department of Inorganic Chemistry, University of Chemistry and Technology Prague, Technick{\'a} 5, 166 28 Prague 6, Czech Republic}
\author{R. Kudrawiec}
\affiliation{Department of Semiconductor Materials Engineering, Faculty of Fundamental Problems of Technology,
 Wroc{\l}aw University of Science and Technology, Wybrze{\.z}e Wyspia{\'n}skiego 27, 50-370 Wroc{\l}aw, Poland}

\begin{document}




\begin{abstract}
Here, we investigate the role of the interlayer magnetic ordering of CrSBr in the framework of $\textit{ab initio}$ calculations and by using optical spectroscopy techniques. These combined studies allow us to unambiguously determine the nature of the optical transitions.  In particular, photoreflectance measurements, sensitive to the direct transitions, have been carried out for the first time. We have demonstrated that optically induced band-to-band transitions visible in optical measurement are remarkably well assigned to the band structure by the momentum matrix elements and energy differences for the magnetic ground state (A-AFM). In addition, our study reveals significant differences in electronic properties for two different interlayer magnetic phases.  When the magnetic ordering of A-AFM to FM is changed, the crucial modification of the band structure reflected in the direct-to-indirect band gap transition and the significant splitting of the conduction bands along the $\Gamma-Z$ direction are obtained. In addition, Raman measurements demonstrate a splitting between the in-plane modes $B^2_{2g}$/$B^2_{3g}$, which is temperature dependent and can be assigned to different interlayer magnetic states, corroborated by the DFT+U study. Moreover, the $B^2_{2g}$ mode has not been experimentally observed before. Finally, our results point out the origin of interlayer magnetism, which can be attributed to electronic rather than structural properties. Our results reveal a new approach for tuning the optical and electronic properties of van der Waals magnets by controlling the interlayer magnetic ordering in adjacent layers.
\end{abstract}
\section{Introduction}
Two-dimensional (2D) magnetic semiconductors give the opportunity to take advantage of the electron charge and the electron spin at the same time, which gives the possibility of broadening modern semiconductor technology and spintronic devices \cite{Awschalom2007} and can greatly expand the applications of ferromagnets (FM) in other devices such as transformers, electromagnets, high-density storage, and magnetic random access memory \cite{Chen2019}. Cr$_2$Ge$_2$Te$_6$ and CrI$_3$ were the first ferromagnetic semiconductors revealed by magneto-optical Kerr microscopy in 2017 \cite{Gong2017,Huang2017}, where the intrinsic long-range FM order was discovered in 2D systems. The study of the layer-number dependence of properties showed that bulk Cr$_2$Ge$_2$Te$_6$ is a Heisenberg ferromagnet, in which the magnetic moments are orientated in all directions, but the few-layer material reveals a good ferromagnetic ordering with excellent gate-voltage tunability \cite{Xing2017}. In the case of CrI$_3$, the number of layers and the stacking arrangement affect the magnetic properties of this material, through the competition between direct Cr-Cr exchange and super-superexchange Cr-I-I-Cr interactions\cite{Sivadas2018,Xu2020}.

CrSBr belongs to ternary chromium chalcogenide halide compounds (CrXh, where X = S/Se and h = Cl/Br/I) that have recently been widely investigated. CrSBr has attracted attention because its bulk, first synthesised 50 years ago, was diagnosed as an antiferromagnetic (AFM) semiconductor \cite{Katscher1966,Goeser1990}, with the bulk N\'{e}el temperature of T$_N$=132$\pm$1~K \cite{Telford2020}. Under ambient conditions, CrSBr has an orthorhombic structure with a P$mmn$ space group. The crystal structure of CrSBr is formed by the two Cr layers bounded by S atoms and passivated by Br layers along the $c$ axes.
CrSBr belongs to van der Waals crystals; therefore, it can be easily exfoliated down to the monolayer using conventional mechanical exfoliation techniques. CrSBr has been identified as an extremely interesting and unique van der Waals type magnetic semiconductor, having the band gap in the near-infrared region between 1.25 eV and 1.5 eV, possessing a large magnetic moment (~3 $\mu$B per Cr atom) \cite{Telford2020,Wilson2021,Yang2021}. CrSBr exhibits an A-type antiferromagnetic order, with robust FM within the layers, and AFM arrangement of adjacent layers (see Fig. S1 in Supporting Information (SI)).

 In this paper, by used of the combined DFT+U studies and experimental measurements  we demonstrate that changes in magnetic phases can be reflected in the optical and electronic  properties of the layered CrSBr.
\section{Experimental and computational details}

  CrSBr was investigated by photoreflectance (PR), optical absorption, photoacoustic spectroscopy (PAS), photoluminescence (PL), and Raman measurements in the 10-300~K temperature range.~For PR experiments the sample was illuminated with a halogen lamp and the reflected light was dispersed with a 0.55~m focal length single grating monochromator.~The signal was detected with a Si photodiode and measured by the lock-in technique. Reflectance was modulated by a chopped (280~Hz) laser beam of a wavelength of 405~nm. Temperature-dependent optical absorption spectra were measured using the same tunable light source as that used for photoreflectance experiments. The intensity of light transmitted through the sample was measured using Si photodiode.~Photoluminescence and RAMAN measurements were made in the microregime, where a 532-nm line of a diode-pumped solid state laser (DPSS) was focused on the sample with a long working distance objective (50$\times$ magnification, NA = 0.55); the diameter of the laser spot was estimated to be below 2 $\mu$m.~PL and RAMAN spectra were measured using a single grating 0.55-m focal length monochromator with a multichannel liquid nitrogen cooled Si CCD array detector.~During these measurements, the sample was mounted on a cold finger in a closed cycle refrigerator coupled with a programmable temperature controller that allows measurements in the 10-300~K temperature range. The photoacoustic spectrum was measured using the gas-microphone method employing an electret condenser microphone mounted in a sealed aluminium cell with a quartz transmission window. The sample was excited with a tunable light source consisting of a 150~W quartz tungsten halogen lamp and a 300 mm focal length monochromator. The excitation beam was then mechanically modulated at a frequency of 40 Hz and focused on the sample surface. More details about the principles of PAS can be found in Refs.~\citenum{Rosencwaig1975, Rosencwaig1978,Zelewski2019}.

  The bulk CrSBr crystal for this study was grown by chemical vapour transport by direct reaction of pure Cr, Br, and S elements in a quartz ampoule. Details of the material synthesis and structural characterisation of CrSBr can be found in SI and Ref.~\citenum{Klein2021}. The magnetic behaviour of the crystal was also tested and found to be antiferromagnetically below 132 K (see the magnetic properties described in SI), in concert with data from the literature \cite{Katscher1966,Goeser1990,Telford2020}.

  DFT calculations have been performed in Vienna Ab Initio Simulation Package (VASP) \cite{KRESSE199615}. The electron-ion interaction was modelled using the projector-augmented wave technique (PAW) \cite{Holzwarth2001}. The Perdew-Burke-Ernzerhof (PBE)\cite{doi:10.1063/1.1926272} exchange-correlation (XC) functional was employed. A plane wave basis cut-off of 550 eV and a $12\times12\times10$ Monkhorst-Pack \cite{Monkhorst1976} k-point grid for BZ integrations were set with  Gaussian smearing of 0.02 eV  for integration in reciprocal space. Non-collinear magnetism and SOC were included in our calculations.
A $1\times1\times2$ unit cell was used to exploit the A-AFM ordering of CrSBr. The semi-empirical D3 Grimme approach \cite{Grimme} was used to account for van der Waals forces.  The position of the atoms and the lattice parameters were optimised within the DFT+U method \cite{PhysRevB.57.1505}. This approach accounts for a proper description of the on-site Coulomb repulsion between 3d electrons of transition-metal ions, by using effective Hubbard U parameters. We have carefully checked the dependence of the U parameter on the electronic bands close to the band gap (see SI). The electronic structure within the DFT + U approach was then compared with the electronic structure obtained within the hybrid functional HSE06 \cite{doi:10.1063/1.2404663}.  U = 3 eV was chosen to calculate the optically active direct transitions and the phonon dispersion curves. The former were determined by calculating direct interband momentum matrix elements using the wave function derivatives within the density functional perturbation theory \cite{PhysRevB.73.045112}. Phonon dispersion curves were obtained within Phonopy software \cite{TOGO20151} using the Parli{\'n}ski-Li-Kawazoe method \cite{PhysRevLett.78.4063}, applied for the 3x3x2 supercell to find interatomic force constants within the harmonic approximation.

\section{Results and analysis}
\begin{figure}
\includegraphics[width=1.0\columnwidth]{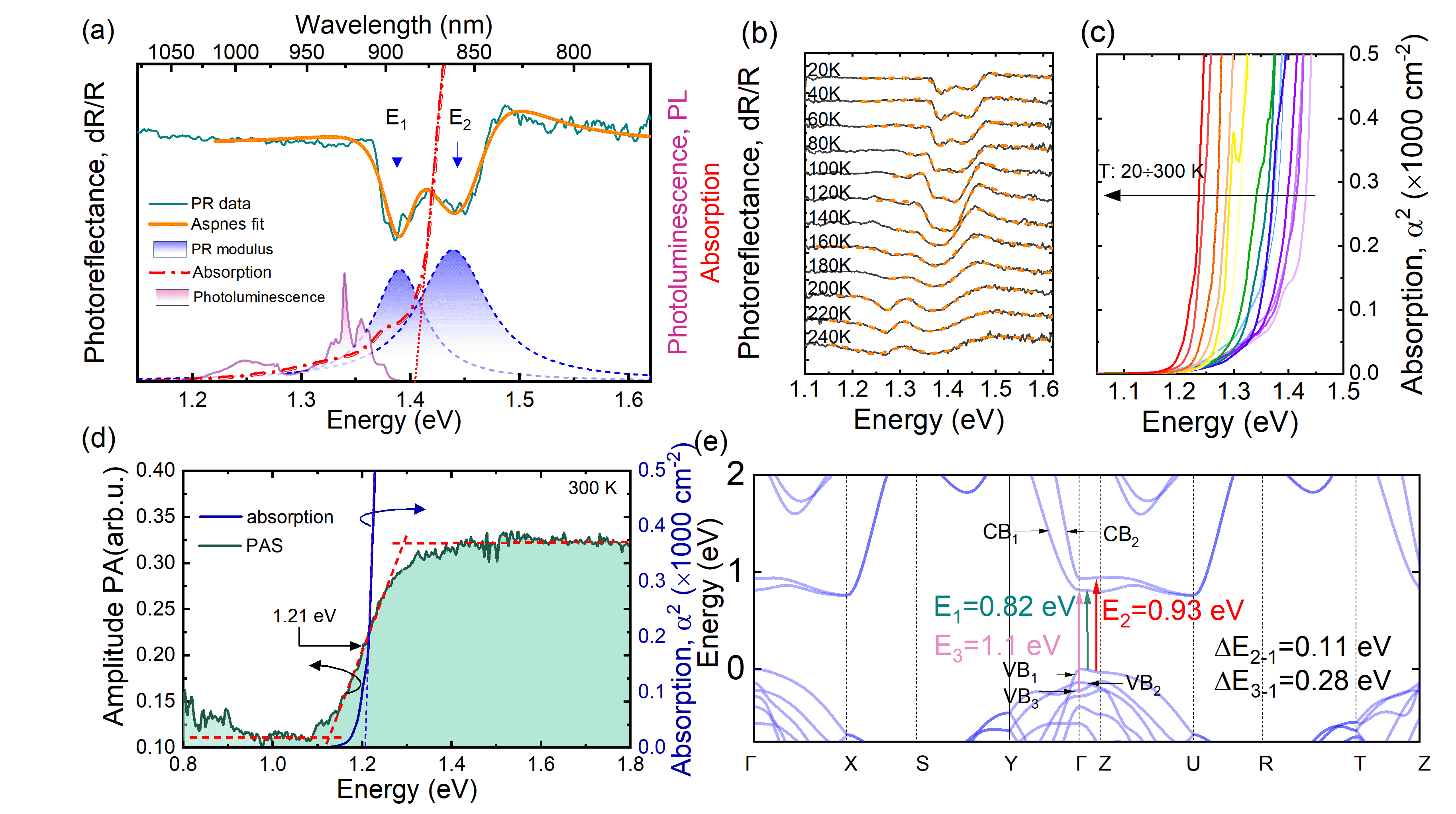}
\caption{(Color online) (a) Photoreflectance, optical absorption, and photoluminescence spectra of CrSBr collected at 20~K.~Aspnes fit to the PR spectrum is also included, and the moduli of the PR transitions are plotted as blue-shaded Lorentzian curves. (b) Temperature dependence of photoreflectance spectra with Aspnes fits (dashed lines) of CrSBr for low-energy optical transitions, (c) temperature-dependent absorption of CrSBr, and (d) comparison of room-temperature optical absorption and photoacoustic spectra. (e) Electronic structure for the magnetic ground state (A-AFM) of bulk CrSBr along high-symmetry lines in the first Brillouin Zone.}
\label{Fig1}
\end{figure}

Figure~\ref{Fig1} presents PR, absorption and PL spectra for a bulk CrSBr acquired at 20~K.~Combining these experimental techniques allows us to unambiguously discern the nature of the optical transitions of CrSBr and establish their energy.~Two resonances, denoted as E$_1$ and E$_2$, can be observed in the PR spectrum. The origin of these optical transitions will be explained later.~Since PR is only sensitive to direct optical transitions and the PR resonances cover the absorption spectrum, we can conclude that all of the measurements presented here reveal the direct character of the optical transition.
~In order to determine the position of the resonance on the energy scale and its broadening, the PR spectrum was fitted using the Aspnes formula.
\begin{equation}
\frac{\Delta{R}}{R}\left(E\right)=\sum_{j}\text{Re}\left[\frac{A_{j}\exp(i\theta_{j})}{(E-E_{j}+i\Gamma_{j})^{m}}\right],
\label{eq_aspnes}
\end{equation}
where $\frac{\Delta{R}}{R}\left(E\right)$ denotes the energy dependence of the PR signal, $A_j$ is an amplitude of the $j$th resonance, $E_j$ indicates the energy of the $j$th optical transition, $\Gamma_j$ is the broadening parameter and $\theta_j$ represents the phase of the resonance line \cite{Aspnes1973}.~The parameter $m$ depends on the character of the optical transition; $m=2.5$ is conventionally used for the band-to-band transition, while $m=2$ is assigned for the excitonic transition.

For a better illustration of the energy transitions, their moduli are plotted in Fig.~\ref{Fig1} (a) below the PR spectrum (see the blue-shaded peaks at the bottom), which are defined as
\begin{equation}
\Delta\rho(E_j)=\frac{|A_j|}{[(E-E{_j})^2+\Gamma_j^2]^{\frac{m}{2}}},
\end{equation}

PL spectrum exhibits several intriguing features. The broad-flat peak at ~1.25 eV we assign to defect-like emission. The doublet at around 1.35 eV comes rather from excitonic emission. Klein $et$ $al.$ have recently suggested that CrSBr behaves like a quasi--1D electronic material manifested by a very narrow exciton and, at low excitation power, appears as a doublet. Such an exciton can inherit the 1D character and show pronounced exciton-phonon coupling effects \cite{Klein2022}.

The temperature dependence of the optical transition of CrSBr has also been comprehensively investigated.~For this purpose PR and absorption measurements have been performed in the temperature range from 20 to 300~K.
Figure~\ref{Fig1} (b) shows the PR spectra and the corresponding fits of CrSBr measured at temperatures between 20 and 300~K in the 1.15-1.65~eV energy range.~The transition energies $E_1$ and $E_2$ at each temperature were determined by fitting the spectra with Eq.~\ref{eq_aspnes}.~It can be observed that the position of the PR resonances moves toward lower energies and broadens as the temperature increases. Furthermore, for temperatures higher than 140~K the shape of the PR spectra drastically differs from those at lower temperatures.
The absorption spectra obtained at different temperatures for CrSBr are shown in Figure~\ref{Fig1}(c).~The absorption onsets vary between 1.21 at 300~K and 1.41~eV at 20~K.~The plots of $\alpha^2$ versus the photon energy exhibit approximately linear behaviour, which is consistent with a direct band gap.~ No features below the sharp onset of $\alpha$ were observed that would be a strong indication of an indirect transition.
Figure \ref{Fig1} (d) shows a comparison of the optical absorption and photoacoustic spectra at room temperature of a bulk CrSBr crystal. Both spectroscopic methods reveal a steep absorption edge in the 1.15-1.3~eV energy range. Note that both methods use monochromatic light illumination with changing wavelengths as opposed to broad-spectrum probing, which can deliver undesirable effects such as excessive sample heating or photovoltaic carrier generation. It is clearly seen that the photoacoustic spectrum follows optical absorption up to 1.3~eV when it saturates, which is a characteristic effect in photoacoustic detection performed at low modulation frequency. PAS does not depend on the analysis of light reflected or passed through the sample and is fundamentally associated with optical absorption. PAS is sensitive to the indirect transition \cite{Zelewski2017} and no evidence of the indirect band gap is observed.

 To provide insight into the observed optical transitions of the bulk phase below T$_N$, the band structure calculations are presented in Fig.~\ref{Fig1} (e) along the high-symmetry lines in the first Brillouin zone. Our electronic band gap obtained for A-AFM within the HSE06 functional (1.85~eV, see SI) is comparable to previous GW calculations for the monolayer (1.8~eV) \cite{Wilson2021}, and experimental measurements using scanning tunnelling spectroscopy and PL (1.5+-0.2~eV) \cite{Telford2020}. The negligible dependence of the Hubbard U on the electronic band gap is predicted (see SI). Although the DFT+U approach underestimates the band gap (0.84~eV), this approach reveals a qualitatively similar electronic band dispersion in the vicinity of the Fermi level as the hybrid functional HSE06 and the previous report on the GW calculations \cite{Wilson2021}. Therefore, this method is chosen for further calculations. The valence band maximum (VBM) is located at the $\Gamma$ point, while the conduction band minimum (CBM) is at the U k-point, resulting in the indirect character of the band gap. Note that the difference between the direct band gap (at the $\Gamma$ point) and the indirect one is just about 20~meV. The bands are double degenerate, in line with the results obtained for the CrSBr bilayer in the corresponding magnetic ground state \cite{Wilson2021}. The $\Gamma$--Z line corresponds to the stacking direction of the layers, thus relatively flat bands (denoted VB$_1$, CB$_1$, CB$_2$ in Fig. ~\ref{Fig1}(e)) are visible in the vicinity of Fermi level, resulting in direct nesting-like transitions. We assign the optical transitions from the VB$_1$ to the CB$_1$, and from the VB$_1$ to the CB$_2$ to respective $E_1$ and $E_2$ resonances, visible in the PR spectrum. Note that the $E_2$ occurs between the $\Gamma$ and the Z point (nesting-like character), for which non-zero dipole matrix elements are computed, reflecting the probability of the transition. Both the $E_1$ and $E_2$ transitions exhibit linear polarisation of the light pointing along the $y$ direction. The theoretical values of $E_1$ and $E_2$ were found to be 0.82~eV and 0.93~eV, respectively, while the corresponding experimental values are higher 1.39~eV and 1.45~eV (at 20~K). Because these absolute values cannot be directly compared between two different approaches, the differences between the transitions $\Delta E_{2-1} = E_2-E_1$ are determined. Namely, $\Delta E_{2-1}$ has been found to be 0.11~eV, which corresponds to the experimental energy separation of 0.07~eV in PR spectra. In the PR we also found the third resonance around 1.77 at 20~K (denoted as $E_3$ -see Fig. S9 in SI), which we assign to the non-zero intensity (see Fig. ~\ref{Fig1}(e)) of the transition from VB$_3$ to CB$_1$ occurring at the $\Gamma$ point. The energy difference between $E_1$ and $E_3$ ($\Delta E_{3-1}$) from PR has been found to be 0.38~eV. This value corresponds to the DFT value of 0.28~eV, which supports our assignment.

\begin{figure}[!h]
\includegraphics[width=0.6\columnwidth]{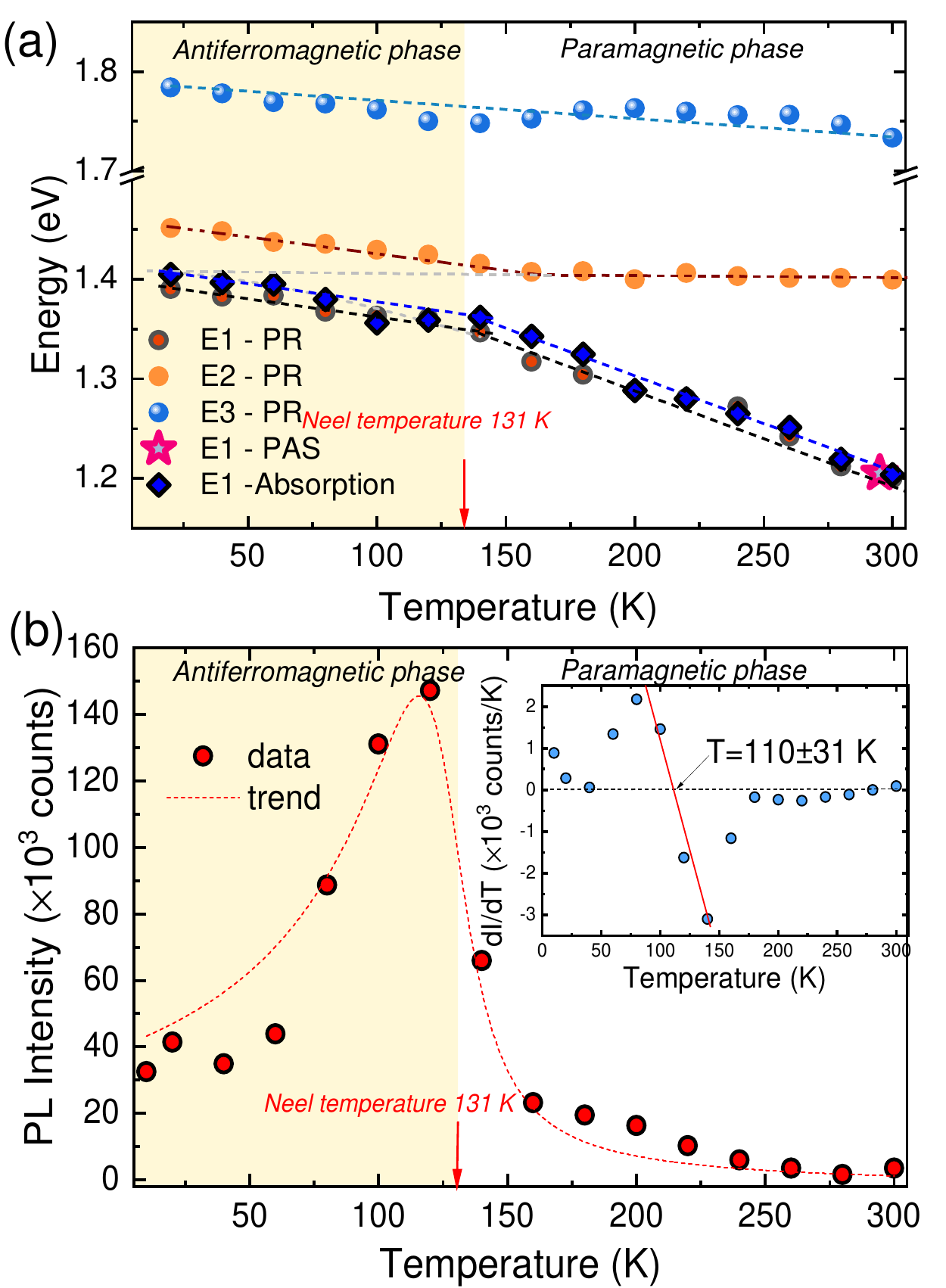}
\caption{(Color online) (a) Energy of optical transitions observed for CrSBr as a function of temperature. The circles are energies of E$_1$ and E$_2$ obtained from PR measurements. Diamonds denote E$_1$ determined by absorption, and a star is E$_1$ as a result of photoacoustic spectroscopy. (b) PL intensity as a function of temperature.}
\label{Fig4}
\end{figure}

Next, we present the temperature dependence of the optical transitions ($E_1$, $E_2$ and $E_3$) in Fig.~\ref{Fig4}(a).~Filled circles denote results obtained from the PR and Aspnes fitting, while diamonds represent the optical absorption results, where the values were found using standard linear extrapolation to the intersection with the background level of the $\alpha^2$ versus $h\nu$ curves. The temperature evolution of the E$_1$ and E$_2$ transitions has a linear trend with a kink around 140~K, resulting in two linear parts of the temperature dependence. For the 20~K-140~K temperature range, the E$_1$ transition redshifts from 1.39 to 1.34~eV in the case of PR results and from 1.41 to 1.36~eV in the case of absorption results, which gives the change of $\sim4.2\times10^{-4}$~eV/K. The redshift of E$_2$ in the same temperature range is from 1.45 to 1.41~eV, corresponding to a reduction of $\sim3.3\times10^{-4}$~eV/K.
For the 140~K-300~K temperature range, the E$_1$ transition energy changes from 1.34~eV to 1.20~eV for PR results and from 1.36~eV to 1.20~eV for absorption results, which corresponds to a higher reduction of $\sim9.1\times10^{-4}$~eV/K. For both spectroscopic methods, E$_1$ is 1.20~eV at 300~K, which is consistent with the PAS result. While the redshift of the E$_1$ transition is clearly visible at the 140~K-300~K temperature range, the evolution of the E$_2$ temperature differs and reaches the plateau, giving a constant value of $\sim$1.40~eV.
The presence of the kink at $\sim$140~K in Fig.~\ref{Fig4}(a) nearly coincides with the N\'{e}el temperature in the investigated CrSBr crystal, suggesting that the magnetic phase transition occurs and can be observed in optical measurements such as PR or absorption. Furthermore, such a phase transition can influence the shape of the PR spectra; as mentioned above, the shape of the PR spectrum for CrSBr changes for temperatures higher than 140~K.

The temperature-dependent intensity of the PL (TDPL) is shown in Fig.~\ref{Fig4}(b). Usually, in paramagnetic semiconductors, the PL intensity monotonously decreases as the temperature increases because of the enhancement of phonon-exciton interactions at high temperatures, and the PL position shows a redshift following the conventional Varshni equation. However, when the temperature increases, the intensity of CrSBr PL gradually increases and reaches the maximum value near the N\'{e}el temperature, while the intensity of PL becomes weaker at higher temperatures. This suggests that a correlation between light emission and antiferromagnetic ordering might exist in CrSBr. To determine the transition temperature from the temperature-dependent PL spectra, the intensity-temperature curve gradient is calculated and shown in the inset of Fig.~\ref{Fig4}(b). The slope-temperature curve shows a linear region of 100 to 150~K. The linear fit can be used to extract the maximum intensity temperature of PL when the slope is zero, which is T =110~$\pm$~33~K. This value agrees well with the N\'{e}el transition temperature T$_N$ = 131~K of the bulk CrSBr. We also noticed that the PL intensity-temperature crossover is very broad, from 100 to 150~K. A similar correlation between PL intensity and N\'{e}el temperature in another antiferromagnetic van der Waals semiconductor MnPS$_3$ was found recently by Zhou $et$ $al.$ \cite{Zhou2022}. 

\begin{figure}[!h]
\includegraphics[width=1.15\columnwidth]{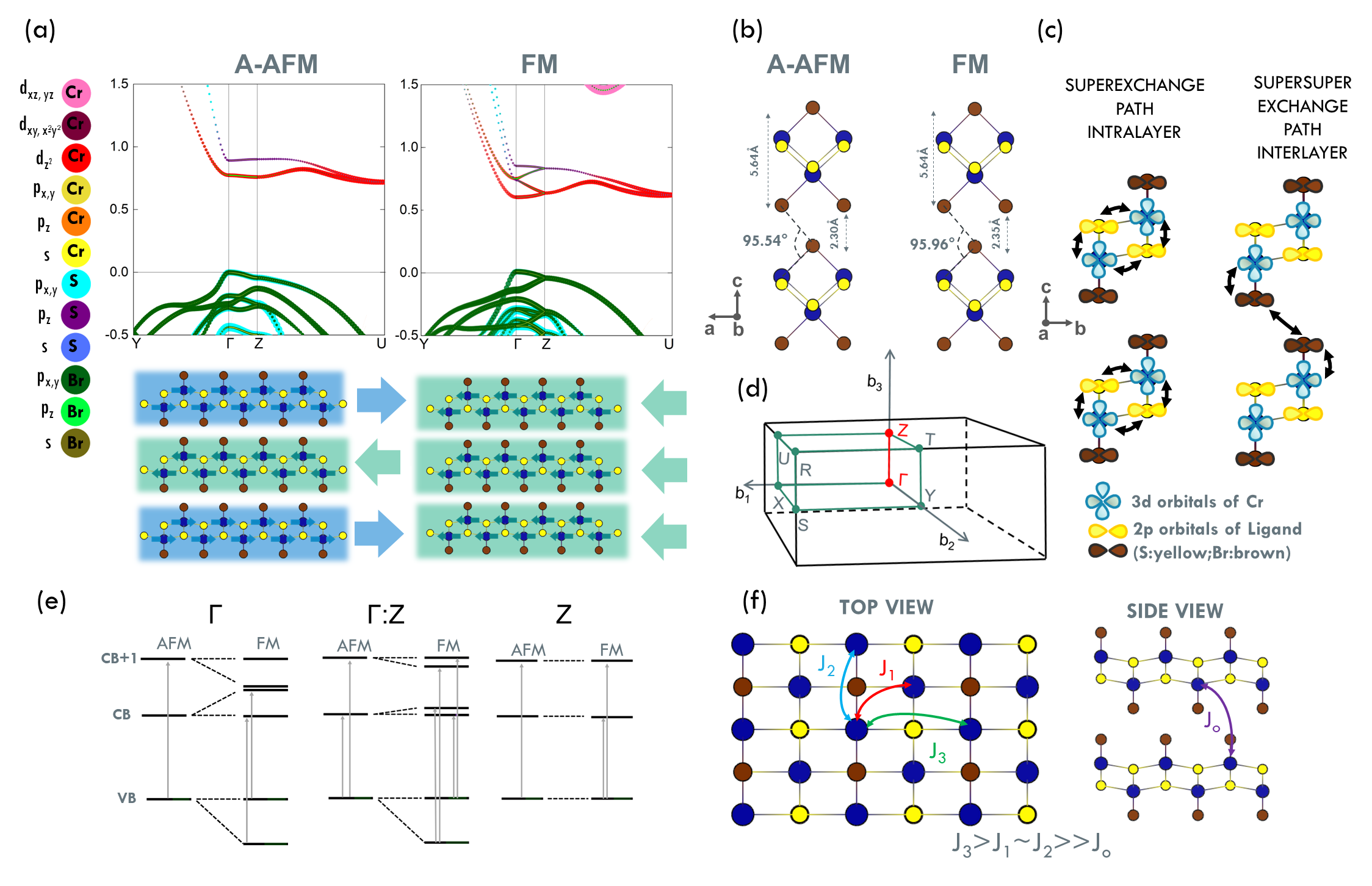}
\caption{(a) Projected band structure of CrSBr for FM and A-AFM magnetic arrangements within the DFT+U approach with a  visualisation  of their corresponding magnetic orderings (bottom of the picture). (b)  Side view of the structural changes of the two adjacent layers in the magnetic arrangements. Schematic presentation of the (c) superexchange and  super-superexchange interactions promoted by the $p$-states of nonmagnetic ligands. (d) The first BZ of the bulk CrSBr along high symmetry directions (denoted in green). (e) Schematic diagrams of the band edges states at particular k-points for both employed magnetic orderings. The arrows indicate the optical transitions with non-zero oscillator strength. All considered transitions were linearly polarised along the $y$ direction. (f) Scheme of the in-plane ($J_1,J_2, J_3$ and out-of plane ($J_o$) magnetic exchange constants.}
\label{Fig5}
\end{figure}


\begin{figure}
\includegraphics[width=1.05\columnwidth]{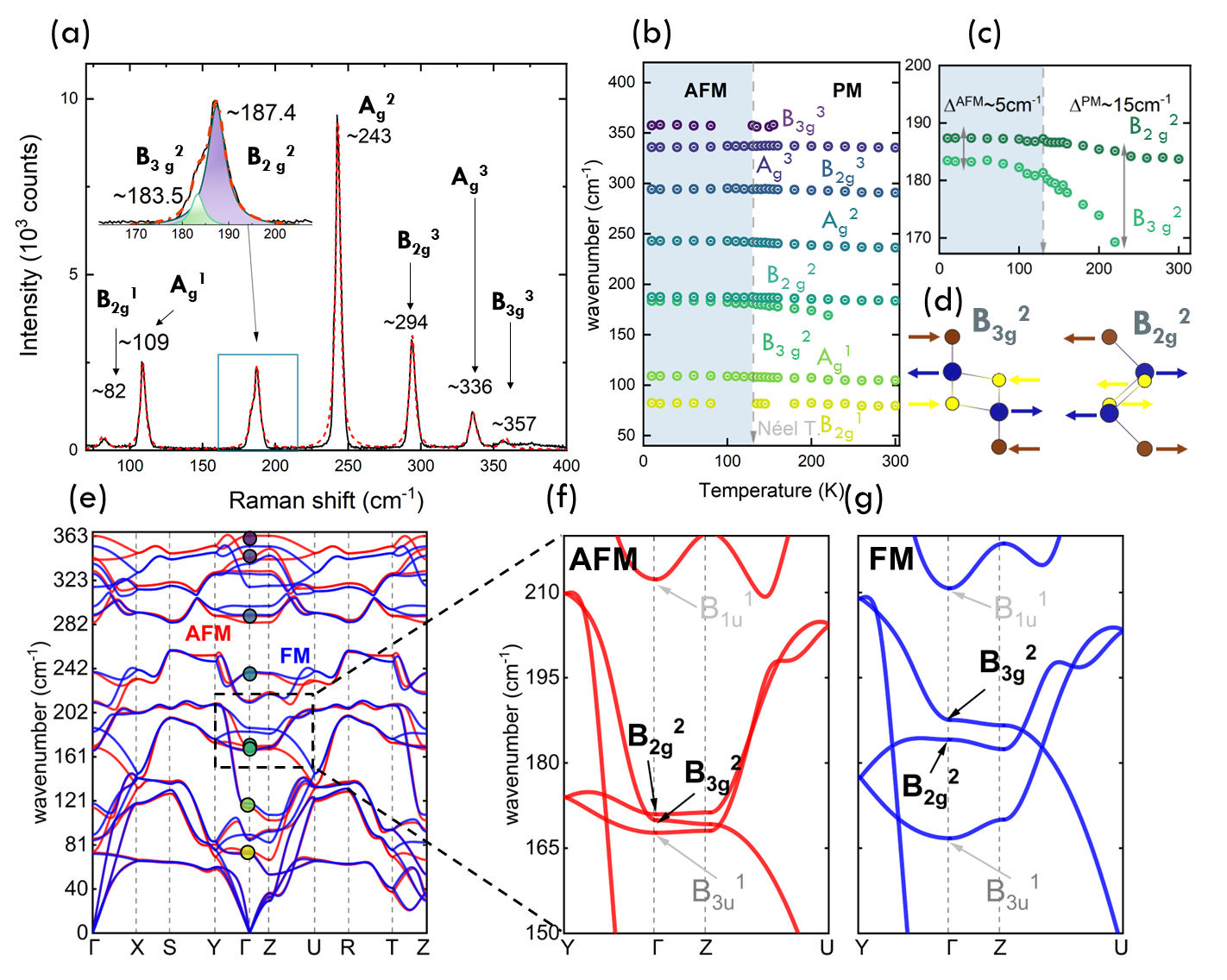}
\caption{(Color online) (a) Raman spectrum for bulk CrSBr obtained at 10K, excited at 532 nm. The wavelengths of the Raman peaks were obtained from Lorentzian fitting. (b) Temperature dependence of the position of all measured Raman peaks. Particular temperature curves for (c) in-plane phonon modes ($B_{2g}^2$, $B_{3g}^2$), and (d) their corresponding schematic representations. (e) Calculated phonon dispersion for two magnetic states (red- AFM, blue-FM) with a  zoom in (f,g)the vicinity of the $\Gamma$ point. }
\centering
\label{Fig6}
\end{figure}

Next, we present the electronic structure assuming two magnetic states FM and A-AFM of bulk systems (see Fig.~\ref{Fig5}(a, b)), along the high-symmetry lines in the IBZ.

Significant splitting of the conduction bands is visible in the FM case, which is absent in A-AFM arrangement along the $\Gamma$-Z direction (see Fig.~\ref{Fig5}(a)), not been reported yet in previous DFT calculations \cite{Wilson2021, Cenker2022}. This aspect directly reflects the difference between the stacking direction of the magnetic arrangements of the employed states, whereas the $\Gamma$ -Z direction corresponds to this stacking direction in the first BZ. First, the direct character of the band gap is obtained for the FM case, and the degeneracy of the bands is lifted compared to that of the A-AFM state. The magnetic ordering breaks the time-reversal symmetry (TRS). However, in the case of an A-AFM arrangement, the out-of-plane translation symmetry shift recovers the symmetry in the system, and the bands remain degenerate. Splitting the bands in the FM case results in a wider band dispersion of the conduction bands compared to the AFM case, and in general depends on the overall magnetisation of the system (see Fig. S10 in SI). A pronounce splitting of the bands is visible for the conduction bands at $\Gamma$, resulting in lowering of the conduction band minimum, and  leading to the change in the character of the band gap from the indirect (A-AFM) to direct one (FM). The band gap is lowered by 110~meV  for FM compared to the A-AFM case. Due to the splitting of the bands, more transitions might occur for the FM case  along the $\Gamma$-Z directions (see Fig.~\ref{Fig5}(e)). 
In addition, optical matrix elements calculated (see SI)  for the transitions around the Fermi level reveal the linear polarisation of the light (y component), irrespective to the magnetic ordering. Note that the dominant contribution to the VBM along the $\Gamma-Z$ direction comes from $p_{x,y}$ orbitals of Br (FM) and S (A-AFM) atoms. There is also a different hybridisation between the $3d_{z^2}$ Cr states  and $p$ state of Br at CBM for the employed magnetic arrangements (FM - $p_{x,y}$, A-AFM - $p_z$).  
These electronic structure changes should be reflected in magnetic interactions for two employed magnetic arrangements.  Note that the $p$ orbitals of both ligands (S, Br atoms) are involved in the superexchange mechanism within the layers, as indicated in Fig.~\ref{Fig5}(c). However, only the $p$ states of Br atoms mediate the magnetic interaction within the super-superexchange mechanism between adjacent layers. Although, it has been shown that in-plane exchanges ($J_1, J_2, J_3$) are one order of magnitude larger than out-of-plane $J_0$ (schematic picture of exchnage couplings in Fig.~\ref{Fig5}(f)), which is AFM and equal to -1.5~meV \cite{Klein2022}, the $J_0$ is important and cannot be neglected because it controls overall magnetisation (see Fig. 10 in SI). 


Next, we address how the interlayer magnetic ordering affects the phonon spectra. For this purpose, we compare the theoretical phonon dispersion for the A-AFM and FM phases, with the experimentally obtained Raman spectrum excited at 532~nm and collected at 10~K (see Fig.~\ref{Fig6} (a)). During the experiment, a sample was tilted with respect to the laser beam - this was forced by the lumpy shape of the sample.
This experimental spectrum shows well-defined peaks (in cm$^{-1}$) at ~82, ~109, ~243, ~294, ~336, and 357 and two overlapping peaks in the vicinity ~188~cm$^{-1}$ (Table 1 with experimental and theoretical Raman modes can be found in SI). This fits the previously reported Raman modes for crystals with P$mmn$ symmetry, where of the 15 phonon modes, only A$_g$, B$_{2g}$, and B$_{3g}$ are Raman-active \cite{Bykov2013,Zhang2019}.
Each peak is fitted by Lorentzian curves. Then we examine the positions of these peaks as a function of temperature (see Fig.~\ref{Fig6} (b)).
We observe strong spin-phonon coupling for the A$_g^3$ (see Fig. S11 (b)), which have been already observed in previous works \cite{}. Here, we detect a remarkable
 splitting for the B$_{2g}^2$/B$_{3g}^2$ (see Fig.~\ref{Fig6} (c,h)) modes, which up to the N\'{e}el temperature have a constant position difference (5$cm^{-1}$) at a temperature close to the  N\'{e}el temperature, start to move away from each other (up to 15~$cm^{-1}$ at 220~K). Our theoretical calculations reveal that the difference between the B$_{2g}^2$ and B$_{3g}^2$ modes depends on the magnetic order (4~$cm^{-1}$ for FM, 1~$cm^{-1}$ for A-AFM), and in the case of FM order the frequency of these modes are blue shifted by 13 $cm^{-1}$ (B$_{2g}^2$) and 18 $cm^{-1}$ (B$_{3g}^2$) compared to A-AFM. 
Furthermore, the significant change in the position of the out-of-plane A$_g^3$ mode is obtained experimentally (see Fig. S11 (b,c)), whereas no difference between the position of this mode is revealed for the FM and A-AFM cases (see SI and Table 1 in SI), excluding the origin in the interlayer magnetism. In fact, our theoretical considerations have corroborated the previous DFT study (Ref.~\citenum{Xu2022}), which has shown that the change in the position of A$_g^3$ is sensitive to the in-plane magnetic order, pointing to the strong correlation between spin and phonon.

Finally, we examine the structural changes between the employed magnetic arrangements and consider the interlayer magnetism from the point of view of its structural or electronic origin. The structural difference between the A-AFM and FM orderings is visible in the distance between the layers and the representative angle between Br-Br-Cr (see Fig.~\ref{Fig5}b)). These quantities are enlarged by 2$\%$ and  0.4$\%$ in comparison to A-AFM case, respectively, while no structural changes are observed for in-plane lattice parameters and in-plane bond lengths. Note that the out-of-plane optical phonon modes A$_g$ are nearly intact upon magnetic order (see Fig.~\ref{Fig6}(g)), suggesting that the interlayer structural changes have negligible impact on the magnetic arrangement. However, the significant shift in energy and separation
 between in-plane B$_{2g}$ and B$_{3g}$ phonon modes at the $\Gamma$ k point are visible. Thus, we claim that the interlayer magnetism is rather of electronic than structural origin.  This result is strongly corroborated by our electronic band structure calculations presented above. Note that there is a significant difference between the states at the top of VB, which are mainly composed of Br $p_{x,y}$ (in FM) and S $p_{x,y}$ (A-AFM) along the $\Gamma-Z$ direction. These states are mainly involved in the superexchange mechanism within the layer (see Fig.~\ref{Fig5}(c)).


\section{Summary}
The results presented here demonstrate that the magnetic phase transition can be manifested in optical spectra. The temperature dependences of the absorption edge and PR resonances do not follow the trend observed in conventional semiconductors (i.e. the Varshni trend). Here, we observe a distinct kink at N\'{e}el temperature, which indicates the change in the magnetic phase of CrSBr. Furthermore, the intensity of PL increases to N\'{e}el temperature and then decreases as the temperature increases. This behaviour suggests a correlation between light emission and antiferromagnetic ordering.
A careful analysis of the band structure shows that the experimental results are very consistent with the DFT calculations. The energy differences between the experimental optical transitions are in good agreement with the energy differences determined theoretically. A significant difference between the interlayer magnetic order is visible in band structure calculations, along the $\Gamma - Z$ direction.  That is, the band gap is lowered by 110~meV, and the character of the band gap changed from indirect (A-AFM) to direct (FM). The remarkable splitting of the conduction bands is obtained for the non-zero overall magnetisation. Moreover, the in-plane phonon modes ($B_{2g}$, $B_{3g}$) are sensitive to the interlayer magnetic ordering, whereas the out-of-plane modes ($A_g$) are intact to the interlayer magnetic arrangement.  In particular, the increase in splitting between $B_{2g}$ and $B_{3g}$ obtained for FM compared to A-AFM correlates with the increase of $B_{2g}$/$B_{3g}$  splitting measured across the N\'{e}el temperature. Therefore, we point out the relevance of other magnetic states, which might be present close to the N\'{e}el temperature, as reported in \citenum{Lee_2021}. Visible in-plane structural changes (interlayer distances and angles) do not affect out-of-plane phonon modes ($A_g$). The significant changes are the electronic properties upon the different interlayer magnetic orderings, pointing out the electronic rather than structural origin of the interlayer magnetism in CrSBr.

\section{Acknowledgements}
This work was carried out under the grant of the National Science Centre, Poland (Grant No. 2019/35/B/ST5/02819).
M.B. acknowledges
support from the University of Warsaw within the project
“Excellence Initiative-Research University” program.
Access to computing facilities of PL-Grid Polish Infrastructure for Supporting Computational Science in the European Research Space, the Wroclaw Centre for Networking and of the Interdisciplinary Center of Modeling (ICM), University of Warsaw are gratefully acknowledged. We acknowledge ACK Cyfronet AGH (Poland) for awarding this project access to the LUMI supercomputer, owned by the EuroHPC Joint Undertaking, hosted by CSC (Finland) and the LUMI consortium through Pl-Grid organization (Poland), under the grant entitled: "Electronic, optical and thermoelectric properties of selected layered materials  and selected heterostructures".
\section{Associated Content}
Supporting Information (SI) Available:

\section{Conflict of Interest}
The authors declare no conflict of interest.


\newpage
\begin{center}
\Large \emph{SUPPORTING INFORMATION}
\end{center}
\section{CrSBr synthesis and structural characterisation}
 Bulk single crystal CrSBr was prepared using a chemical vapour transport method. Chromium, sulphur and bromine elements with a stoichiometry of 1:1:1 were added and sealed in a quartz tube under high vacuum and then placed in a two-zone tube furnace. The material was pre-reacted in an ampoule at 700~$^{\circ}$C for 10 hours, and the source and growth ends were kept at 850 and 900~$^{\circ}$C, respectively. After 25 h, the temperature gradient was reversed and the hot end gradually increased from 880 to 950~$^{\circ}$C in 5 days period. The high-quality CrSBr single crystals were removed from the ampule in an Ar glovebox.

CrSBr exhibits an A-type antiferromagnetic order, with robust FM within the layers, and AFM arrangement of adjacent layers (see Fig.~\ref{S1}).
\renewcommand{\thefigure}{S1}
\begin{figure}[!h]
\includegraphics[width=0.75\columnwidth]{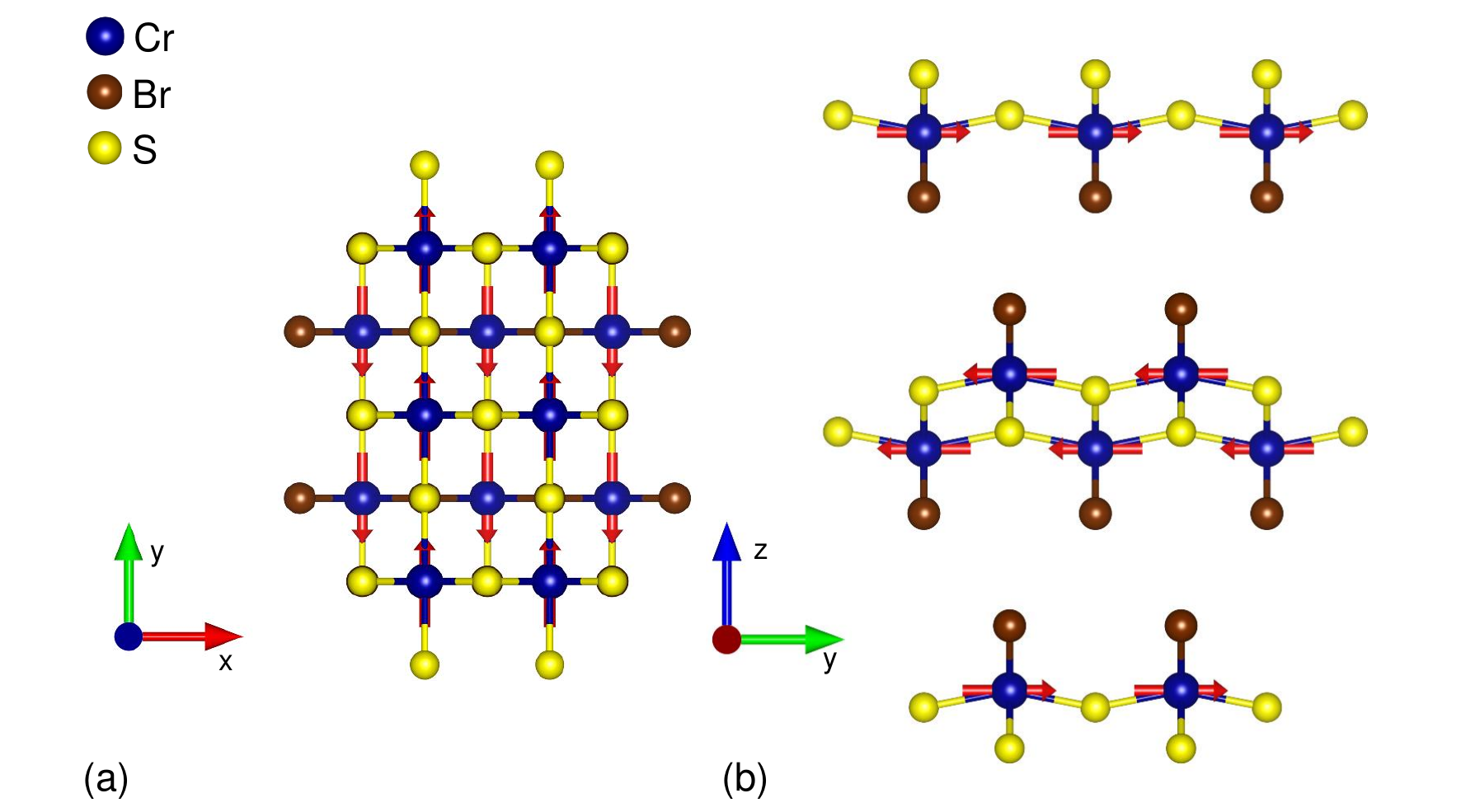}
\caption{(Color online) Different crystal orientations for CrSBr.~The (a)[001], (b) [100] directions are pointing out of the page for the space group $Pmmn$. The magnetic ground state of the CrSBr bulk system (A-AFM) is marked by red arrows. FM alignment of Cr spins within the layer and AFM alignment of Cr spins in adjacent layers.~In panel (b) the layered structure can be clearly seen and the van der Waals gaps between these layers can be distinguished.}
\label{S1}
\end{figure}

 The bulk crystal of CrSBr was characterised using XRD (see Fig. \ref{S2}). Powder XRD data were collected at room temperature on a Bruker D8 Discoverer powder diffractometer (Bruker, Germany) with Parafocusing Bragg-Brentano geometry using CuK$_\alpha$ radiation ($\lambda$=0.15418~nm, U = 40~kV, I = 40~mA). The diffraction patterns were collected at room temperature for 2$\theta$ values from 5-90$^\circ$. The acquired data were analysed using HighScore Plus 3.0 software. XRD reveals a pure single-phase CrSBr with a high preferential orientation because of the layered van der Waals structure.
\renewcommand{\thefigure}{S2}
\begin{figure}
\includegraphics[width=0.6\columnwidth]{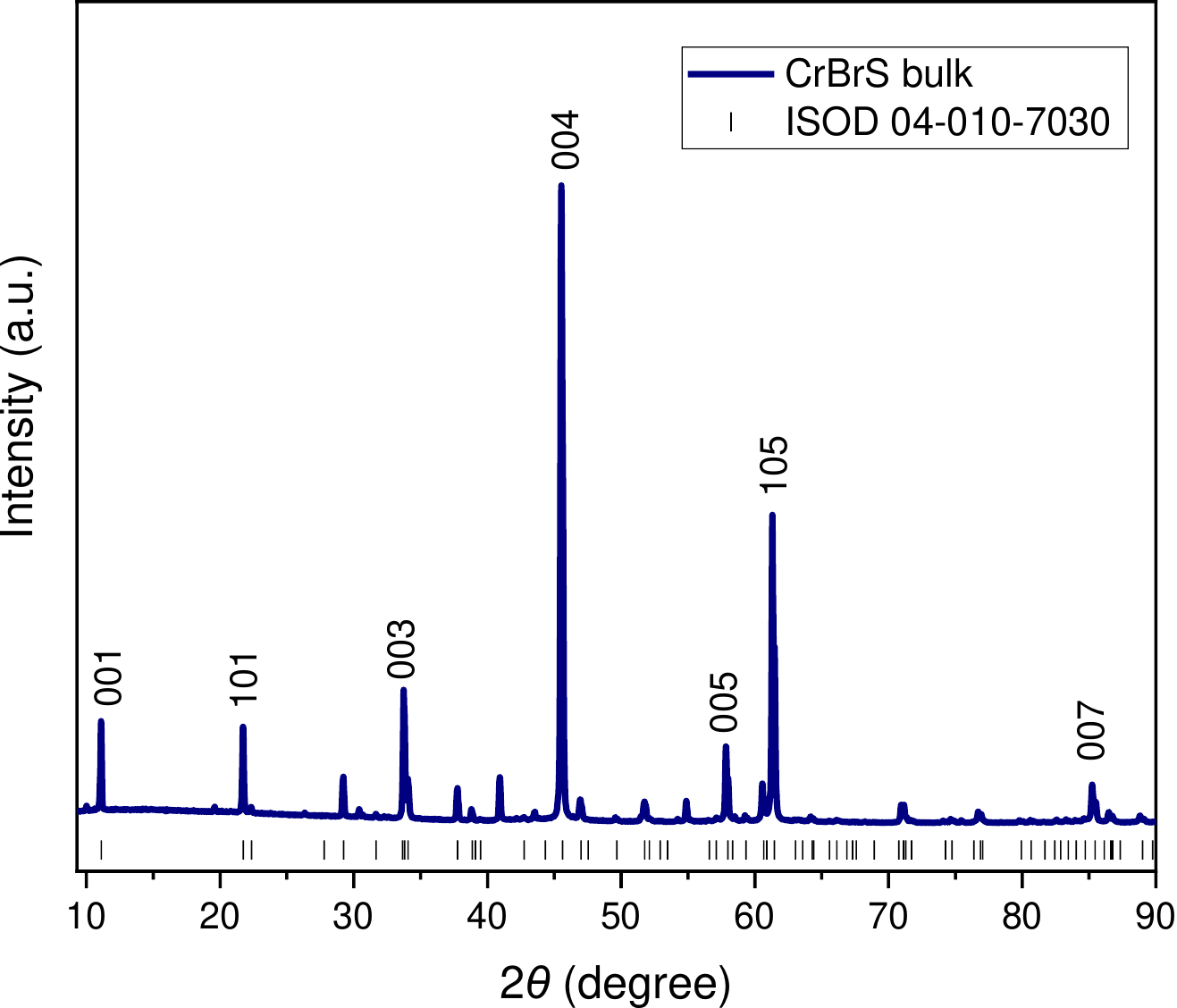}
\caption{(Color online) X-ray diffractogram for bulk CrSBr.}
\label{S2}
\end{figure}

High resolution transmission electron microscopy (HR-TEM) was performed using an EFTEM Jeol 2200 FS microscope (Jeol, Japan). A 200 keV acceleration voltage was used for measurement. Elemental maps and EDS spectra were acquired with SDD detector X-MaxN 80 TS from Oxford Instruments (England). The sample preparation was achieved by drop casting the suspension (1 mg ml$^{-1}$ in acetonitrile) on a TEM grid (Cu; 200 mesh; Formvar/carbon) and dried at 60 $^\circ$C. The morphology was also investigated using scanning electron microscopy (SEM) with a FEG electron source (Tescan Lyra dual beam microscope). The sample was placed directly on carbon tape and measurements were carried out using a 10~kV electron beam. TEM and SEM results are presented in Figure \ref{S3}.
\renewcommand{\thefigure}{S3}
\begin{figure}
\includegraphics[width=1\columnwidth]{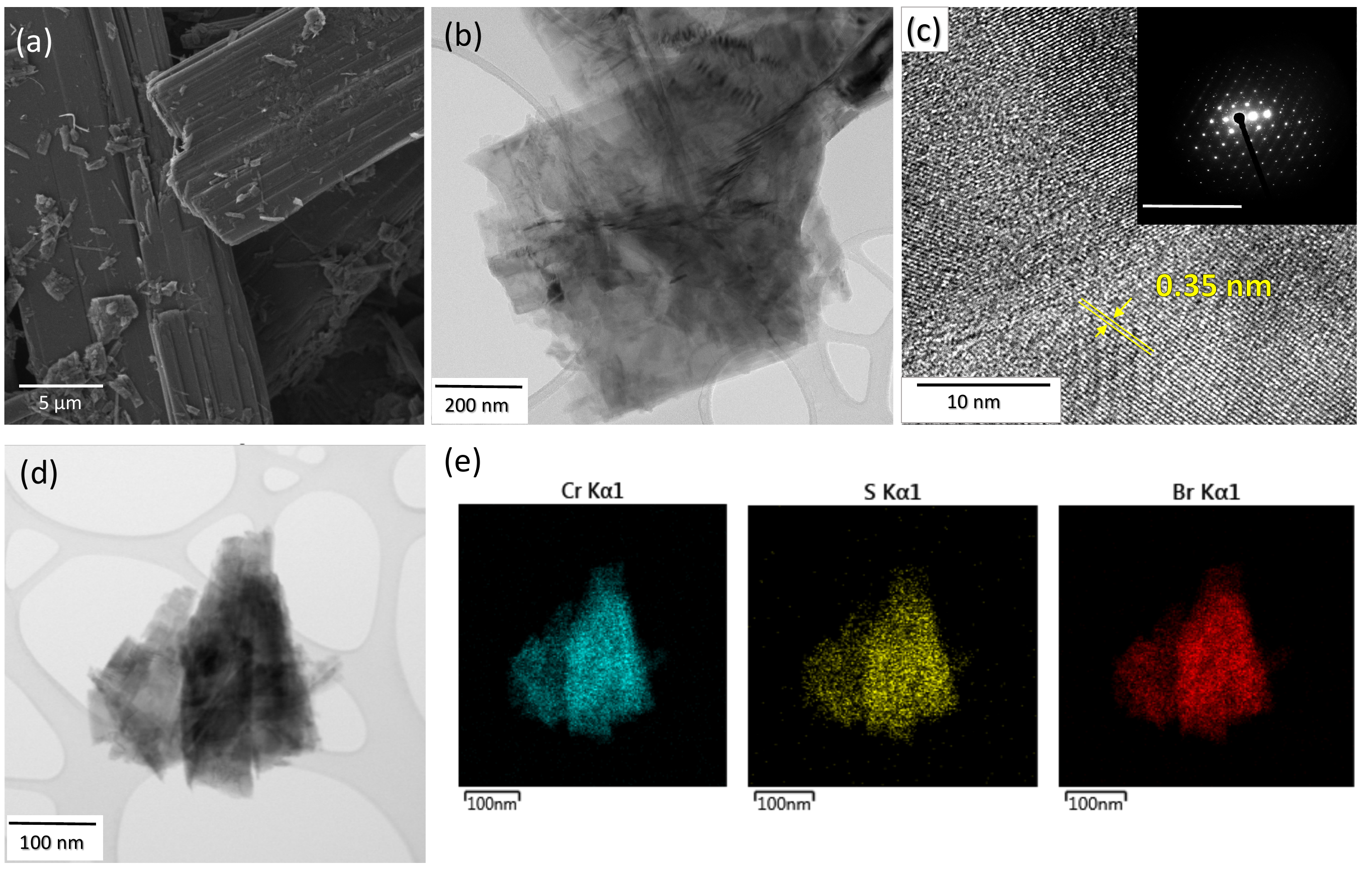}
\caption{(Color online) Morphology and microstructure of CrSBr: (a) Scanning electron micrograph of the bulk. (b,d) TEM image of multilayered CrSBr. (c) High-resolution TEM (HR-TEM) image of (b) sheet surface profile and selected area electron diffraction (SAED), shown in inset with a scale 1/0.04 nm. Yellow mark shows lattice fringes with a d-spacing of 0.35$\pm$0.03 nm. (e) Elemental mapping images of Cr, Br, S.}
\label{S3}
\end{figure}

\newpage
High-resolution X-ray photoelectron spectroscopy (XPS) was performed using an ESCAProbeP spectrometer (Omicron Nanotechnology Ltd, Germany) with a monochromatic aluminium X-ray radiation source (1486.7~eV). Wide-scan surveys of all elements were performed, with subsequent high-resolution scans (Fig.~\ref{S4}). The samples were placed on a conductive carrier made from a high-purity silver bar. An electron gun was used to eliminate sample charging during measurement (1-5~V).

\renewcommand{\thefigure}{S4}
\begin{figure}
\includegraphics[width=1\columnwidth]{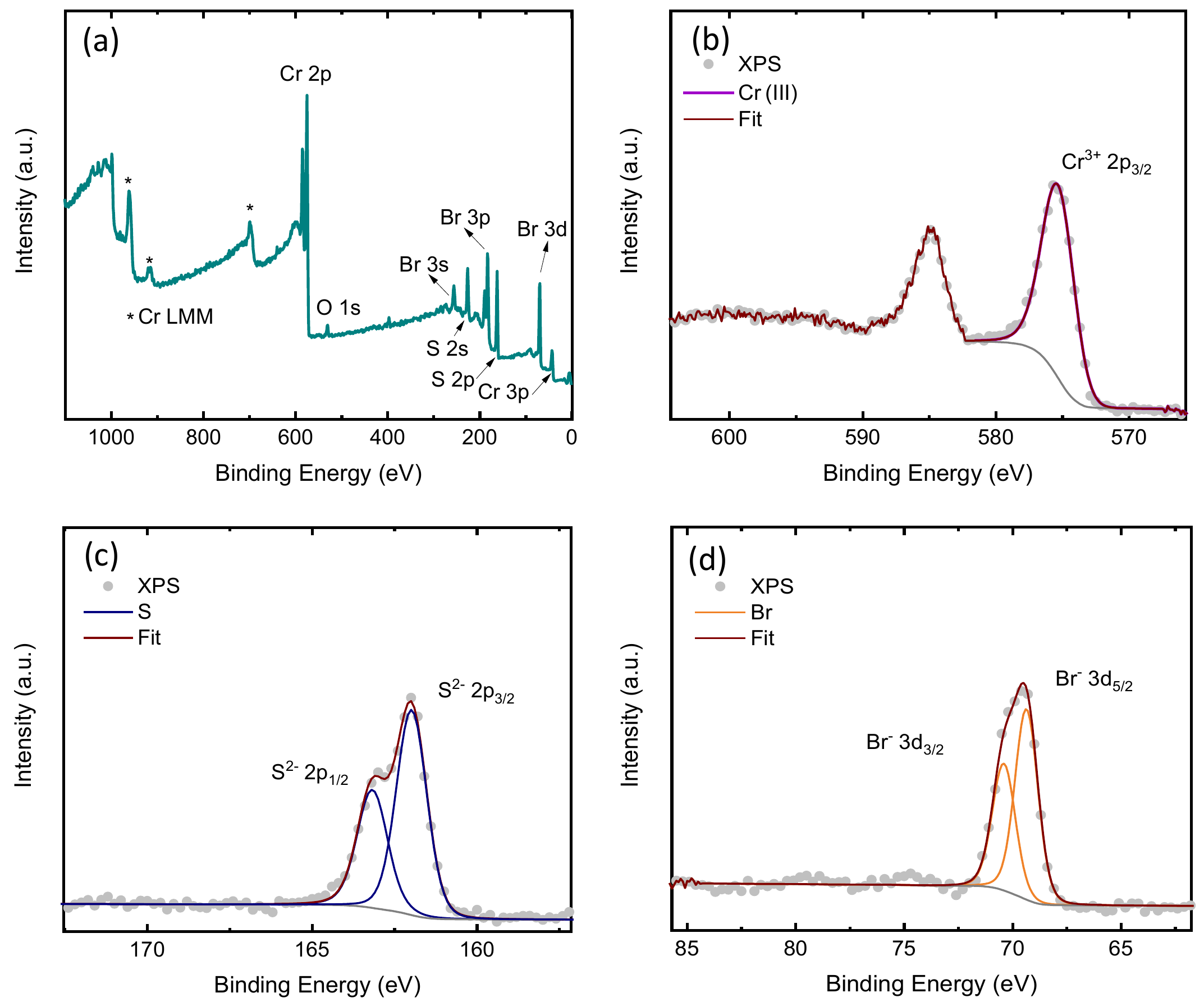}
\caption{(Color online) XPS survey spectrum of the spattered sample (a) and high-resolution spectra of the (b) Cr 2p (c) S 2p and (d) Br 3d states of CrSBr,}
\label{S4}
\end{figure}

\section{Magnetic Properties}

Magnetic properties were investigated in the temperature range of 1.73-300~K and in magnetic fields up to 7~T using a Quantum Design MPMS-XL superconducting quantum interference device (SQUID) magnetometer. Measurements were made on a single crystalline specimen of CrSBr with a magnetic field (H) orientated along the crystallographic direction [001].
\renewcommand{\thefigure}{S5}  
\begin{figure}
\includegraphics[width=1\columnwidth]{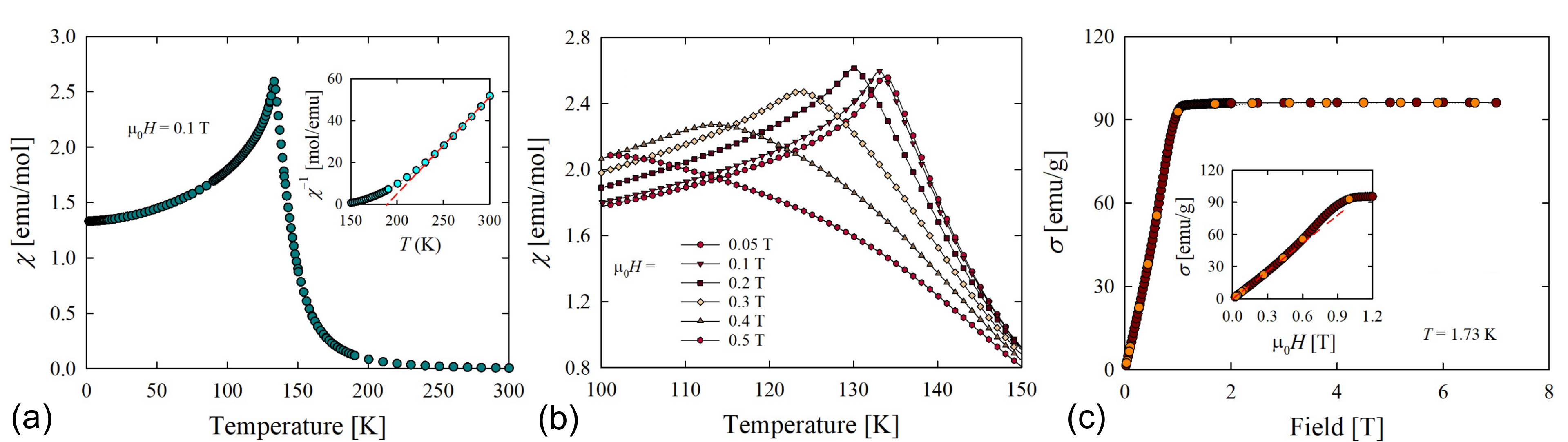}
\caption{(Color online) (a) Temperature variation of the molar magnetic susceptibility of single-crystalline CrSBr measured in a magnetic field of 0.1~T applied along the crystallographic $c$-axis after cooling the sample in a zero field. The inset shows the temperature dependence of the reciprocal magnetic susceptibility. The dashed straight line represents the Curie-Weiss law. (b) Temperature variations of the molar magnetic susceptibility of single-crystalline CrSBr measured as in panel (a) in various external magnetic fields in the vicinity of the AFM phase transition. (c) Magnetic field dependences of magnetisation in single crystalline CrSBr measured at 1.73~K magnetic field applied along the crystallographic $c$-axis. The dark and light symbols represent the data taken with an increasing and decreasing field, respectively. The inset presents the same magnetisation data in small magnetic fields. The dashed line emphasises the initial straight-line behaviour.}
\label{S5}
\end{figure}
As shown in Fig.~\ref{S5}(a), the temperature dependence of the magnetic susceptibility ($\chi$) of CrSBr exhibits a pronounced maximum at T = 132(1)~K that identifies the N\'eel temperature, in conjunction with data from the literature \cite{Goeser1990,Telford2020,Telford2022}. Above about 240 K, $\chi(T)$ follows a Curie-Weiss law (see the inset of Fig.~\ref{S5}(a)) with the effective magnetic moment $\mu{_{eff}}$ = 4.14(5)~µB and the paramagnetic Curie temperature $\theta_p$ = 189(3)~K. Both parameters are similar to those previously reported for the H $\|$$c$-axis \cite{Telford2020}. The experimental value of $\mu{_{eff}}$ is somewhat larger than the effective magnetic moment predicted for a free Cr$^{3+}$ ion with spin-only momentum S = 3/2 (3.87~$\mu{_{B}}$). The large positive value of $\theta_p$ reflects strong ferromagnetic interactions in the plane a-b of the CrSBr crystallographic unit cell \cite{Goeser1990,Telford2020,Telford2022}. Below 240~K, $\chi^{-1}$(T) deviates from a straight-line behaviour, probably due to a crystalline electric field.

In the antiferromagnetic region (AFM), magnetic susceptibility decreases smoothly with decreasing temperature down to 1.73~K, in contrast to data in the literature that indicated an additional anomaly near 35~K of unknown origin \cite{Telford2020,Telford2022}. The latter feature led to a ferromagnetic bifurcation of the $\chi(T)$ curves taken after cooling the specimen in a zero and finite magnetic field, regardless of its direction with respect to the crystal axes \cite{Telford2022}. Remarkably, no such effect was found for the CrSBr crystals investigated in this work, thus proving the purely AFM character of their electronic ground state.

Another clue to the AFM ordering in CrSBr came from the measurements of $\chi(T)$ in various magnetic fields, the results of which are shown in Fig.~\ref{S5}(b). With increasing field strength, the susceptibility maximum systematically shifts toward a lower temperature and broadens, in a manner characteristic of antiferromagnets. As can be inferred from the figure, in a magnetic field of 0.4~T, $\chi(T)$ forms an extended maximum near 112~K, but in $\mu_{0}H$ = 0.5~T, $\chi(T)$ remains featureless down to 100~K.

The latter finding suggests the occurrence of an order-order magnetic phase transition below 0.5 T, which was observed in the magnetisation ($\sigma$) data measured as a function of the magnetic field. As can be seen in the inset of Fig.~\ref{S5}(c), at T = 1.73~K, $\sigma$ is initially a linear function of $H$; however, near 0.45~K, the $\sigma(H)$ isotherm shows an inflection typical for a metamagnetic change in the spin structure. Another anomaly in the isotherm $\sigma(H)$, in the form of a sharp kink, occurs near the critical field $\mu_{0}H_c$ = 0.9~T. The latter feature can be attributed to the formation of a field-forced ferromagnetic alignment of all spins along the direction of the magnetic field. The magnitude of $H_c$ is very similar to the value reported in the literature for the H$\|$$c$-axis \cite{Telford2020,Telford2022}. In strong magnetic fields, the magnetisation saturates at a value of 96(1)~emu/g that corresponds to the magnetic moment $\mu{_{sat}}$ = 2.82(2)~µB that is fairly close to the theoretical value of a free Cr$^{3+}$ ion (3 $\mu{B}$). As shown in Fig.~\ref{S5}(c), the $\sigma(H)$ isotherm measured at T = 1.73~K does not exhibit a hysteresis effect, corroborating the AFM nature of the magnetic ordering in the bulk CrSBr crystals investigated.

\section{Photoreflectance}
Figure \ref{S6} shows temperature-dependent PR for the E$_3$ transition of CrSBr.
\renewcommand{\thefigure}{S6}
\begin{figure}[h]
\includegraphics[width=0.45\columnwidth]{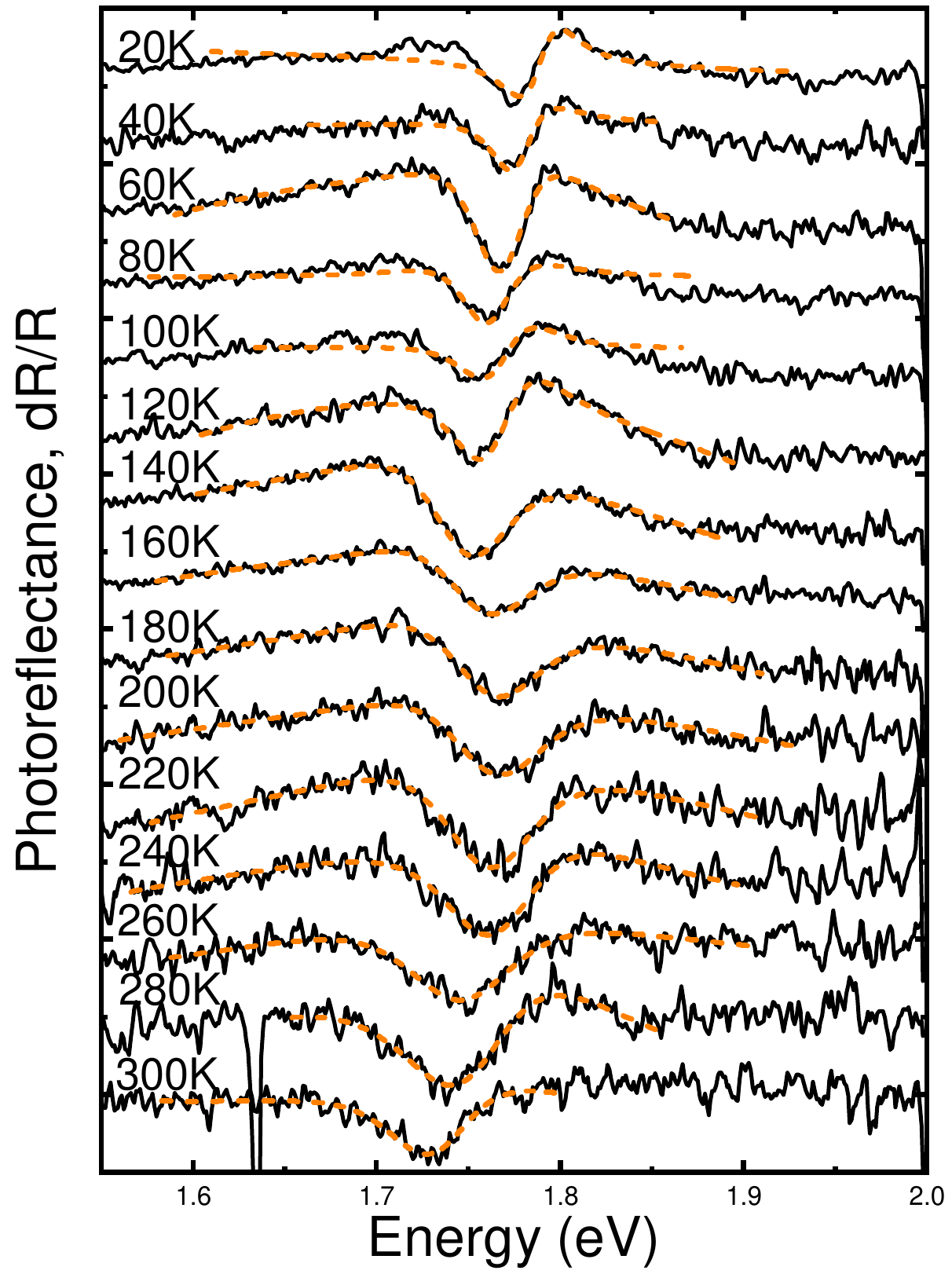}
\caption{(Color online) Temperature dependence of the photoreflectance spectra of the E$_3$ transition with Aspnes fits (dashed lines) of CrSBr.}
\label{S6}
\end{figure}


\section{DFT calculations}

Figure \ref{S9} shows matrix elements in the Brillouin zone, which denote the probabilities of optical transitions. By comparison with the experimental results, we can conclude that the optical transitions occur in the $\Gamma$--Z path, where the smallest direct energy gap is observed in the DFT results. We assign the transition between the first valence band (VB$_1$) and the first conduction band (CB$_1$) to the resonance $E_1$ in the PR spectrum. This transition takes place between the $\Gamma$ and the Z point, where the bands are flat; therefore, it has rather a nesting-like character. Additionally, according to DFT, such a transition is possible because the matrix element in the $\Gamma$--Z region has a non-zero value.

\renewcommand{\thefigure}{S7}
\begin{figure}
\includegraphics[width=1\columnwidth]{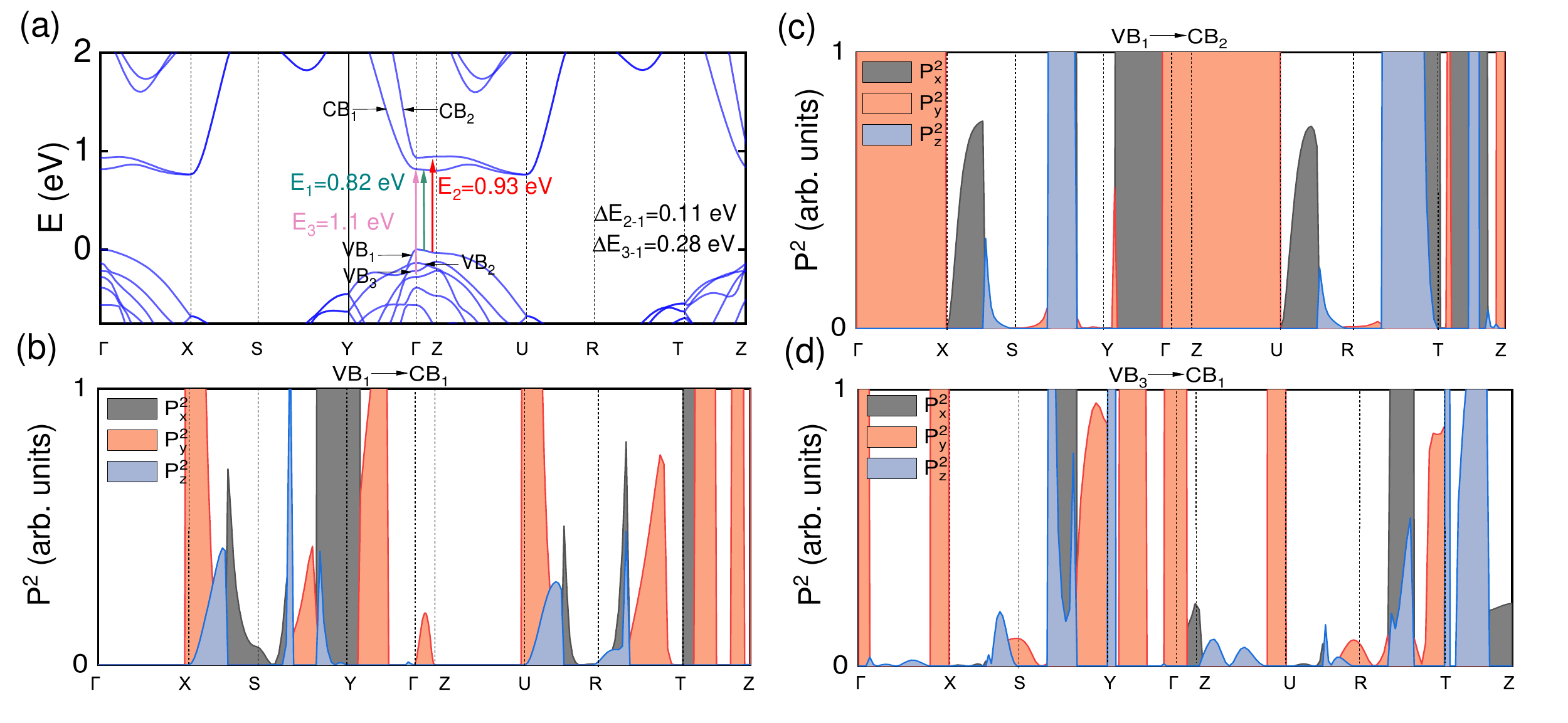}
\caption{(Color online) Bandstruture for the magnetic ground state (A-AFM) in the vicinity of Fermi level along high-symmetry lines. Three optically active transitions are denoted by the green (VB$_1$ to CB$_1$ at the middle of the $\Gamma$-Z line), red (VB$_1$ to CB$_2$ at $Z$ k-point), and pink (VB$_3$ to CB$_1$ at the $\Gamma$ k-point). Note that these transitions exhibit non-zero oscillator strength along the "y" direction. The oscillator strength of the three particular transitions along high-symmetry lines in the first BZ: (c) VB$_1$ to CB$_2$, (d)VB$_1$ to CB$_1$, and (e) VB$_3$ to CB$_1$ are presented.}
\label{S9}
\end{figure}

Figure \ref{S10} presents the comparison of the DFT+U calculation for U = 3 eV and U = 5 eV. No significant differences are observed close to the band gap; therefore, for the analysis the model with U = 3 eV was used. In both cases, the band gap was found to be 0.82 eV.
\renewcommand{\thefigure}{S8}
\begin{figure}[h]
\includegraphics[width=0.8\columnwidth]{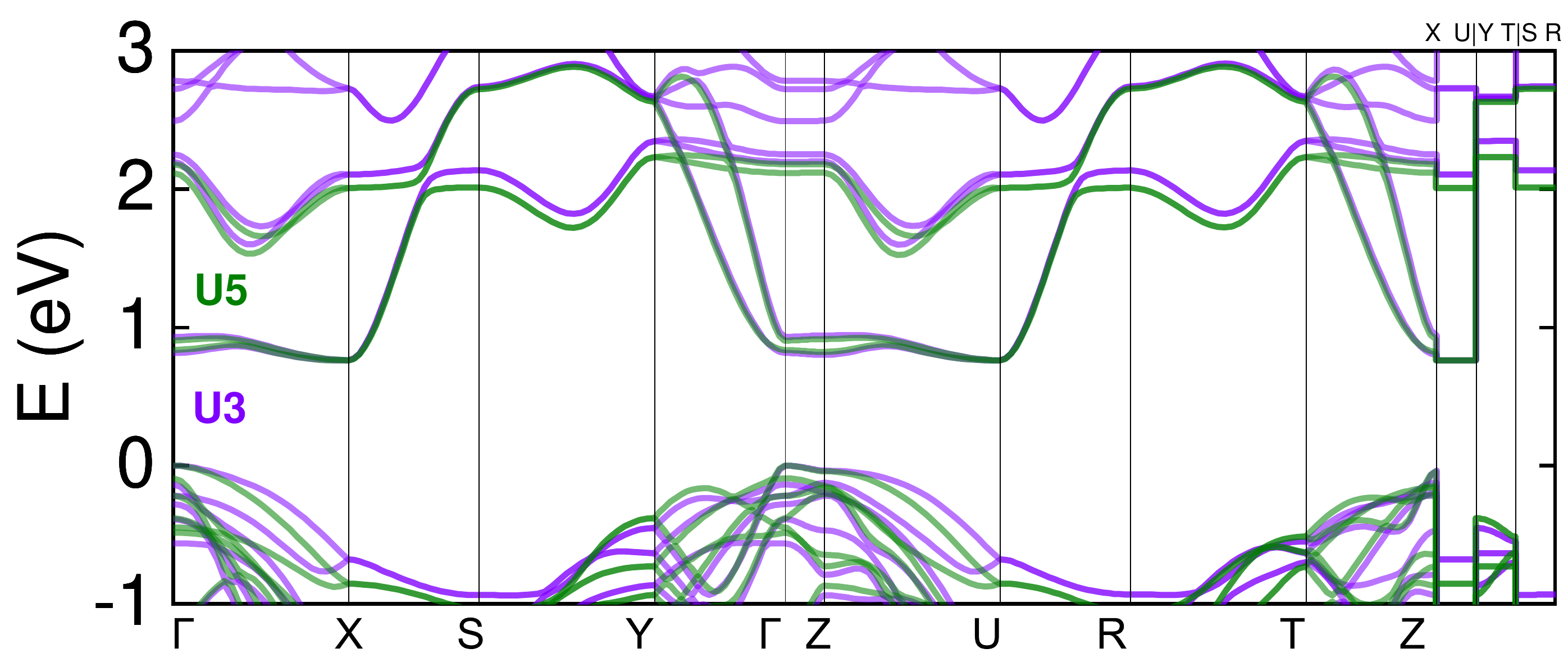}
\caption{(Color online) Comparison of the calculated band structures of CrSBr within the DFT + U framework, employing U = 3 eV (violet line) and U = 5 eV (green line).}
\label{S10}
\end{figure}

Figure \ref{S11} shows the band structure calculated with the hybrid functional HSE06 for bulk CrSBr. In this case, the band gap was found to be 1.85 eV, but the energy differences of $E_2$ and $E_1$, and $E_3$ and $E_1$ are similar to those determined in the framework of DFT+U.
\renewcommand{\thefigure}{S9}
\begin{figure}[h]
\includegraphics[width=0.8\columnwidth]{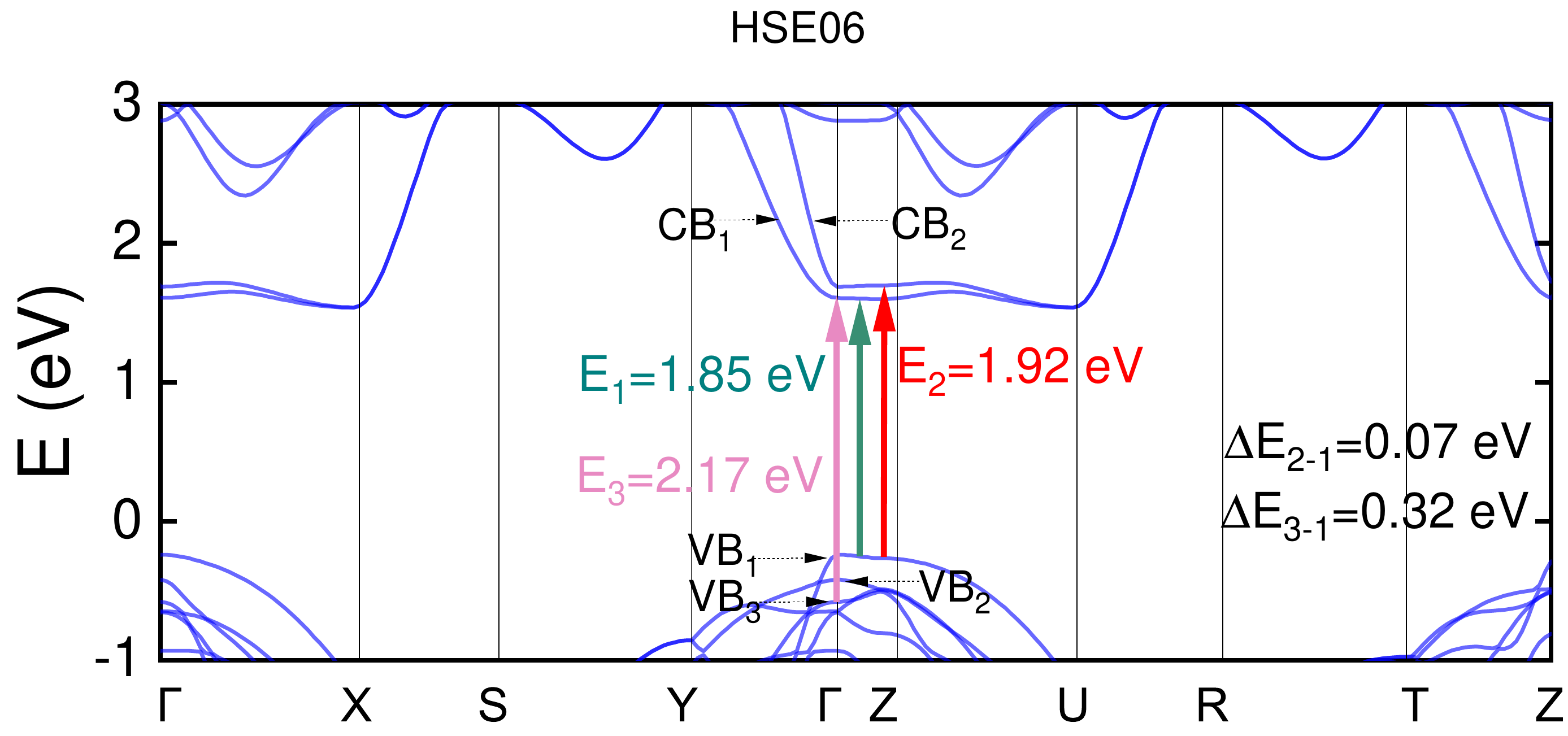}
\caption{(Color online) The band structure of CrSBr was calculated using the hybrid functional HSE06.}
\label{S11}
\end{figure}
\vspace{5cm}
The splitting band dependency on the overall magnetisation in the CrSBr system is shown in Fig.~\ref{S12}.

\renewcommand{\thefigure}{S10}
\begin{figure}
\includegraphics[width=1\columnwidth]{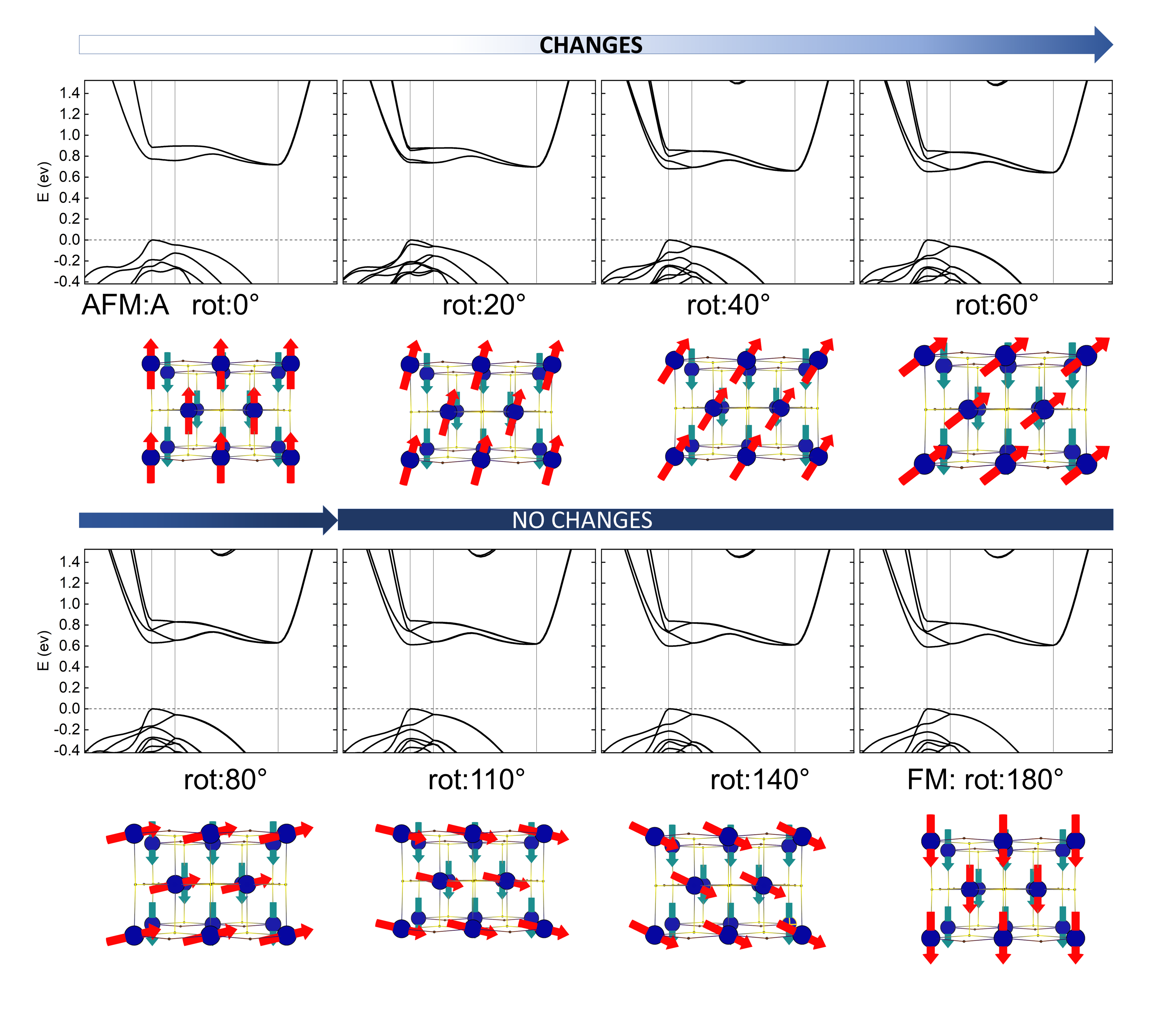}
\caption{(Color online) Band splitting around the $\Gamma$ point upon rotation of the spins within one layer (red arrows) with respect to the fixed alignment of the spins in the adjacent layer (green arrows) as represented by the schematic diagrams below (perspective top view). The particular angle rot:20 indicates the angle between the spins coming from the nearest-neighbour atoms from adjacent layers.}
\label{S12}
\end{figure}

\newpage
\section{Phonons}
\renewcommand{\thefigure}{S11}
\begin{figure}
\includegraphics[width=0.8\columnwidth]{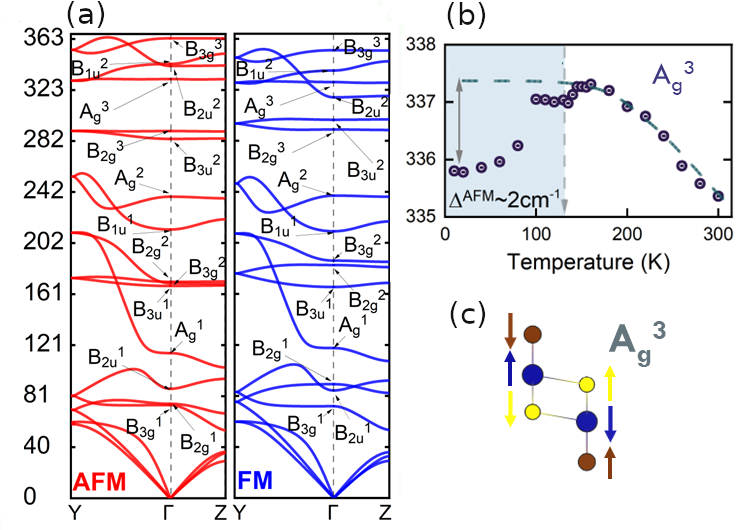}
\caption{(Color online) (a) Calculated phonon dispersion for
two magnetic states, zoom in the vicinity of the $\Gamma$ point. (b) Temperature curves for out-of-plane ($A^3_g$) mode and (c) its corresponding schematic representation.}
\label{S13}
\end{figure}

\begin{table}[]
\centering
\caption{Positions of specific phonon modes for calculations and Raman spectroscopy [$cm^{-1}$].}
\begin{tabular}{l|lll}
                                    & \textbf{AFM} & \textbf{FM} & \textbf{exp} \\ \hline
$\mathbf{B_{3g}^1}$                        & 74           & 73          &              \\
$\mathbf{B_{2g}^1}$                        & 75           & 90          & 82           \\
$\mathbf{B_{2u}^1}$                        & 86           & 85          &              \\
$\mathbf{A_{g}^1}$                         & 115          & 119         & 109          \\
$\mathbf{B_{3u}^1}$                       & 168          & 167         &              \\
$\mathbf{B_{3g}^2}$                        & 170          & 188         & 184          \\
$\mathbf{B_{2g}^2}$                        & 171          & 184         & 188          \\
$\mathbf{B_{1u}^1}$                        & 212          & 211         &              \\
$\mathbf{A_{g}^2}$                         & 238          & 239         & 243          \\
$\mathbf{B_{3u}^2}$                        & 284          & 299         &              \\
$\mathbf{B_{2g}^3}$                        & 290          & 291         & 294          \\
$\mathbf{A_{g}^2}$                         & 331          & 329         & 336          \\
$\mathbf{B_{2u}^2}$                        & 342          & 317         &              \\
$\mathbf{B_{1u}^2}$                        & 343          & 338         &              \\
$\mathbf{B_{3g}^3}$                        & 363          & 354         & 357
\end{tabular}
\end{table}

\newpage
\bibliography{CrBrS2.bib}
\end{document}